\documentclass[twocolumn,aps,prl,amsmath,showpacs,amssymb,floatfix]{revtex4}
\usepackage{tabularx}
\usepackage{bm}
\usepackage{subfigure}
\usepackage{euscript}
\usepackage{graphicx}
\usepackage{color}
\usepackage[colorlinks=true,linkcolor=blue]{hyperref}%
\usepackage{amsfonts}
\usepackage{exscale}
\usepackage{amsbsy}

\begin{document}

\title{Dirac Point Structure in a Bose-Einstein Condensate in a Honeycomb Optical Lattice}
\author{Zhongbo Yan$^1$}
\author{Xiaosen Yang$^2$}
\author{Shaolong Wan$^1$}
\email[]{slwan@ustc.edu.cn} \affiliation{$^{1}$Institute for Theoretical
Physics and Department of Modern Physics University of Science and
Technology of China, Hefei, 230026, P. R. China\\
$^{2}$Beijing Computational Science Research Center, Beijing,
100084, P. R. China}
\date{\today}

\begin{abstract}
We study the Bose-Einstein condensate in a honeycomb optical
lattice within Bogoliubov theory and find that for a ${\bf k} = 0$
condensate, the Dirac points appear in the Bogoliubov excitation
spectrum when $0 < \beta < 2$, which illustrates that the
bose-bose interaction does not change the Dirac point structure
but only give a  modification of the velocity of the Dirac cone.
When the bosons are driven to condense at ${\bf k} = {\bf K}$,
however, we find that the topology of the Dirac points will be
altered by arbitrary weak interaction. Furthermore, we find that
the next-nearest-neighbor hopping in an isotropic and an
anisotropic lattice has different effects to the dynamics of the
condensate and it should be taken into account when the lattice is
not sufficiently deep.
\end{abstract}


\maketitle

In recent years, the Dirac point structure has arisen a lot of
interests in condensed matter physics\cite{A. H. Castro Neto, M.
Z. Hasan, X. L. Qi}, non-linear optics\cite{O. Peleg,M. J.
Ablowitz,O. B. Treidel} and cold atomic physics\cite{L. H.
Haddad1,L. H. Haddad2}. The Dirac point structure corresponds to
many important phenomena in physics, such as the room-temperature
quantum Hall effect in graphene\cite{K. S. Novoselov}, topological
edge state in topological insulator\cite{M. Konig}, and conical
diffraction in honeycomb photonic lattices\cite{O. Peleg}. To
realize such a structure in a cold atomic system, several groups
proposed to load fermions on different optical lattices\cite{S. L.
Zhu,K. L. Lee,Jing-Min Hou,R. Shen} and recently, in experiment,
L. Tarruell et al has realized it in a tunable honeycomb
lattice\cite{Leticia Tarruell}. Before this realization of Dirac
point structure with Fermi gas, the honeycomb lattice was first
investigated with  Bose-Einstein condensates\cite{Parvis
Soltan-Panahi1, Parvis Soltan-Panahi2}, although new quantum
phases were observed, no signatures of Dirac points were observed.

In this paper, we study the Bose condensates in a honeycomb
optical lattice and find that in the tight-banding limit, for a
${\bf k}=0$ condensate, stable Dirac points appear in the two
lowest bands even in the presence of the interaction if the
anisotropy $\beta$ (the meaning of $\beta$ is given in the
following) is within the region $(0,2)$. However, if the bosons
are condensed at ${\bf k}={\bf K}$, where ${\bf K}$ is the
momentum where the two bands touch, even within Bogoliubov theory,
we find that no matter how weak the interaction is, the  Dirac
points will be altered by the interaction, which agrees with the
conclusions obtained in Refs.\cite{O. B. Treidel,Zhu Chen} by
numerical methods. Furthermore, we find that for a lattice that is
isotropic but not sufficiently deep, the next-nearest-neighbor
(NNN) hopping effectively affects the dynamics of the condensate
but nearly does not affect the dynamics around the Dirac points.
However, for a anisotropic lattice, the NNN hopping will alter the
topology of the Dirac points and make the Dirac points no longer
well defined. Therefore, to observe the Dirac points, it's better
to use an isotropic lattice.

$Model$-The honeycomb optical lattice, which consists of two
sublattices A and B, can be realized by three detuned
standing-wave lasers, with the optical potential given by\cite{S.
L. Zhu}
\begin{eqnarray}
V(x,y)=\sum_{i=1,2,3}V_{i}\sin^{2}
\left[k_{L}(x\cos\theta_{i}+y\sin\theta_{i})+\pi/2 \right],
\nonumber
\end{eqnarray}
where $\theta_{1}=\pi/3$, $\theta_{2}=2\pi/3$, $\theta_{3}=0$, and
$k_{L}$ is the optical wave vector. With different $V_{i}$, the
honeycomb lattice can be either isotropic or
anisotropic\cite{L.-M. Duan}. In this work, we consider
single-component bosonic atoms in this lattice. For bosons, the
intra-species collisions is dominated by s-wave scattering. In the
following, we consider $r_{0} << k^{-1} a_{s}$, where $r_{0}$ is
the effective interaction length, $k = \frac{2\pi}{\lambda}$ is
the wave vector, and $a_{s}$ is the s-wave scattering length.
Furthermore, we first assume the lattice is deep, therefore, the
effective Hamiltonian is given as the Bose-Hubbard model
\begin{eqnarray}
H&=&-\sum_{<ij>}\left(t_{ij} \hat{a}_{i}^{\dag} \hat{b}_{j} + h.c.
\right) - \sum_{i\in A} \mu \hat{a}_{i}^{\dag} \hat{a}_{i} -
\sum_{i\in B} \mu \hat{b}_{i}^{\dag} \hat{b}_{i} \nonumber \\
&&+\frac{U}{2} \left[ \sum_{i\in A} \hat{n}_{i}^{a}
\hat{n}_{i}^{a} + \sum_{i\in B} \hat{n}_{i}^{b} \hat{n}_{i}^{b}
\right], \label{1}
\end{eqnarray}
where $<ij>$ represents the nearest neighbor (NN) sites,
$\hat{a}_{i}$ and $\hat{b}_{i}$ denote the bosonic  mode operators
for the sublattices A and B, respectively. $\mu$ is the chemical
potential and $U$ describe the on-site interaction between bosons.
The tunnelling rates $t_{ij}$, in general, depend on the
tunnelling directions in an anisotropic honeycomb lattice. In this
paper, both the isotropic case and the anisotropic case are
analyzed in detail.

First, we make a Fourier transformation for the Hamiltonian. With
$\hat{a}_{i}^{\dag}=(1/\sqrt{N}) \sum_{\bf{k}} exp(i\bf{k} \cdot
\bf{A_{i}})$, $\hat{b}_{i}^{\dag} = (1/\sqrt{N}) \sum_{\bf{k}}
exp(i\bf{k} \cdot \bf{B_{i}}) \hat{b}_{{\bf k}}^{\dag}$, where $N$
is the number of sites of the sublattice A (or B). For a strong
optical lattice, the restriction to the lowest Bloch band is
justified, and the Hamiltonian can be given as
\begin{eqnarray}
H&=&\sum_{{\bf k}} \left[\phi ({\bf k})\hat{a}_{{\bf k}}^{\dag}
\hat{b}_{{\bf k}} + h.c. - \mu
\hat{a}_{{\bf k}}^{\dag} \hat{a}_{{\bf k}} - \mu \hat{b}_{{\bf k}}^{\dag} \hat{b}_{{\bf k}} \right] \nonumber \\
&&+\frac{U}{2N} \sum_{{\bf q}} \left[\hat{\rho}_{{\bf q}}^{a}
\hat{\rho}_{-{\bf q}}^{a} + \hat {\rho}_{{\bf q}}^{b}
\hat{\rho}_{-{\bf q}}^{b} \right], \label{2}
\end{eqnarray}
where $\hat{\rho}_{{\bf q}}^{\alpha} = \sum_{{\bf p}}
\hat{\alpha}_{{\bf p-q}}^{\dag} \hat{\alpha}_{{\bf p}}$ $(\alpha =
a, b)$. $\phi({\bf k}) = \sum_{s=1}^{3} t_{s} exp(i\bf{k} \cdot
\bf{b_{s}})$, with $\bf{b_{1}} = (1/\sqrt{3},1)(a/2)$, $\bf{b_{2}}
= (1/\sqrt{3},-1)(a/2)$, and $\bf{b_{3}} = (-a/\sqrt{3},0)$, $a$
is the lattice spacing. In this paper, we set $t_{1} = t_{2} = t$
and $t_{3} = \beta t$. Furthermore, we set $t$ and $a$ as the
energy unit and length unit, respectively.

For the free case, i.e. $U = 0$, this Hamiltonian can be rewritten
in a $2 \times 2$ matrix form,
\begin{eqnarray}
H = \sum_{{\bf k}}(\hat{a}_{{\bf k}}^{\dag}, \hat{b}_{{\bf k}}^{\dag})
\left(\begin{array}{cc}
- \mu & \phi({\bf k}) \\
\phi^{*}({\bf k}) & - \mu \\
\end{array}\right)
\left(\begin{array}{c} \hat{a}_{{\bf k}} \\
\hat{b}_{{\bf k}}
\end{array}\right), \nonumber
\end{eqnarray}
by making a transformation $(\hat{\alpha}_{{\bf k}},
\hat{\beta}_{{\bf k}})^{T} = V (\hat{a}_{{\bf k}}, \hat{b}_{{\bf
k}})^{T}$, where the matrix $V$ takes the form
\begin{eqnarray}
V = \frac{1}{\sqrt{2}|\phi({\bf k})|} \left( \begin{array}{cc}
|\phi({\bf k})| & - \phi({\bf k}) \\
-\phi^{*}({\bf k}) & |\phi({\bf k})| \\
\end{array}\right), \label{3}
\end{eqnarray}
the Hamiltonian is diagonalized and the energy eigenvalues are
\begin{eqnarray}
\varepsilon_{\pm}(k) = - \mu \pm |\phi({\bf k})|. \label{4}
\end{eqnarray}
Since the bosons are condensed into the zero crystal momentum,
$\varepsilon_{-}(0)$ should take zero, and therefore $\mu = -
|\phi(0)|$. Such a non-zero value of chemical potential is common
in lattice models and it guarantees the energy of excitations to
be positive, which is necessary for the condensates to be stable.

\begin{figure}[bp]
\subfigure{\includegraphics[width=4cm,height=3cm]{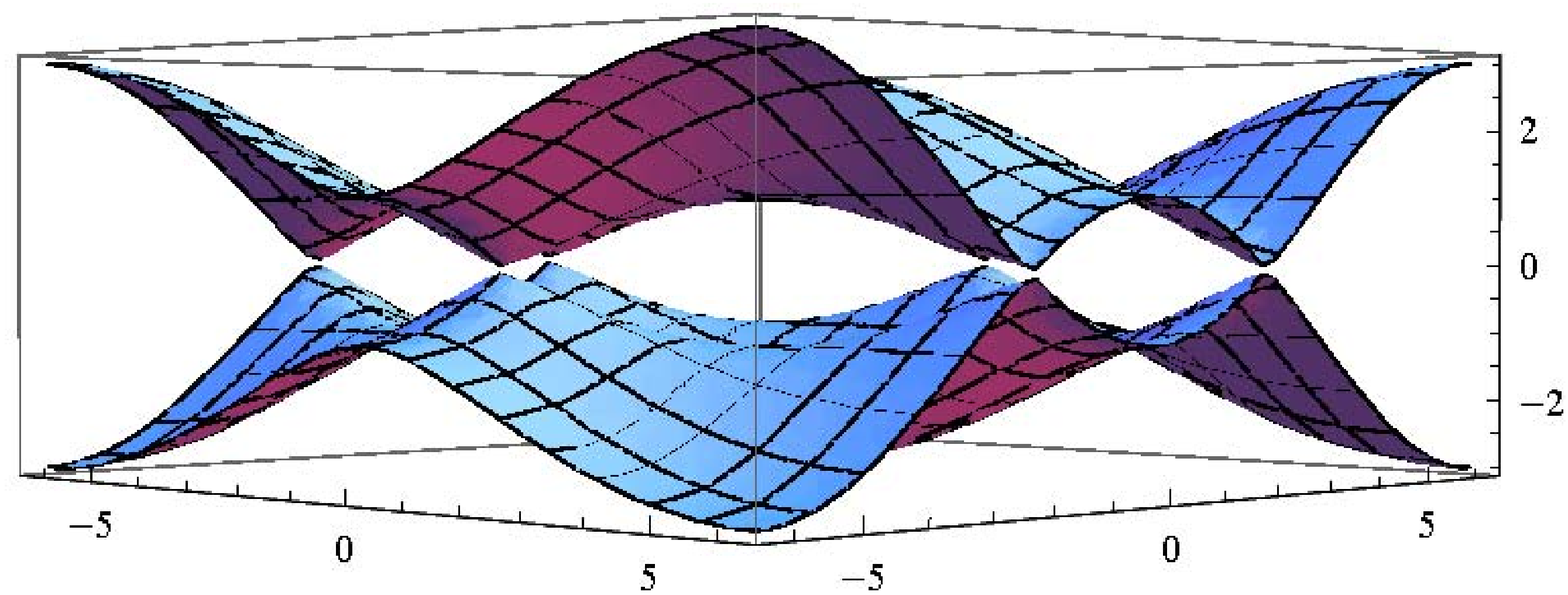}}
\subfigure{\includegraphics[width=4cm,height=3cm]{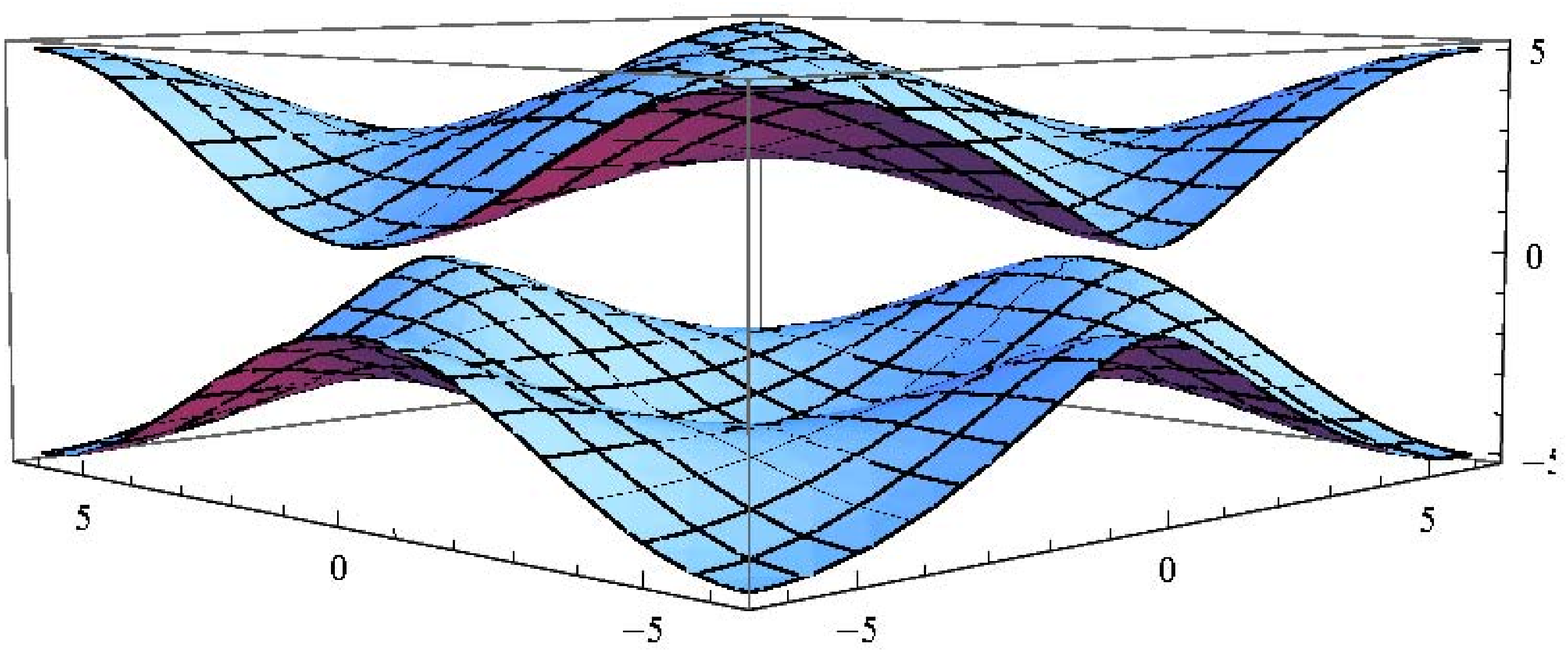}}
\caption{(Color online) (Right) the isotropic case, i.e.
$\beta=1$. The two bands touch at six points, with two different
kind of  Dirac points. (Left) the anisotropic case with $\beta=3$,
the bands are gapped. \label{Fig.1}}
\end{figure}

The dispersion relations are determined by $|\phi({\bf k})|$,
which is directly related to the hopping amplitude and takes the form
\begin{eqnarray}
|\phi(k)| = \sqrt{2 + \beta^{2} + 2 \cos(k_{y}) + 4 \beta \cos(\frac{\sqrt{3} k_{x}}{2}) \cos(\frac{k_{y}}{2})}. \nonumber \\
\label{5}
\end{eqnarray}
From Eqs.(\ref{4}), (\ref{5}) and Fig.\ref{Fig.1}, we see, for $0
< \beta < 2$, the two bands touch at several points and Dirac
points appears. In particular for the isotropic case where $\beta
= $, the dispersion has the same Dirac structure as the graphene
material (as shown in Fig.\ref{Fig.1}). The Dirac points are
purely due to the structure of the honeycomb lattice, without any
relation to the quantum statistics. For $\beta > 2$, the two bands
no longer touch with each other and they are gapped.

When the interaction is tuned on, the chemical potential has a
shift: $\mu = - |\phi(0)| \rightarrow - |\phi(0)| + U n$, where
$n$ is the density of the system. In the Bogoliubov approximation
and neglecting the constant condensation energy, the Hamiltonian
can be written as $H_{BG} = \frac{1}{2} \sum_{{\bf k}} B_{{\bf
k}}^{\dag} H_{{\bf k}} B_{{\bf k}}$, where $B_{{\bf k}}^{\dag} =
(a_{{\bf k}}^{\dag}, a_{-{\bf k}} ,b_{{\bf k}}^{\dag}, b_{-{\bf
k}})$ and
\begin{eqnarray}
H_{{\bf k}} = \left(\begin{array}{cccc} \varepsilon_{0}+U & U& \phi({\bf k}) & 0 \\
U & \varepsilon_{0}+U& 0 & \phi({\bf k}) \\
\phi^{*}({\bf k}) & 0  & \varepsilon_{0}+U & U  \\
0 & \phi^{*}({\bf k}) & U   & \varepsilon_{0}+U \end{array}\right). \nonumber
\end{eqnarray}
Here we assume the filling factor of bosons is unit and
$\varepsilon_{0} = |\phi(0)|$ just for convenience. Based on this
Hamiltonian, we find that the dispersion relations are
\begin{eqnarray}
\varepsilon_{\pm}({\bf k}) = \sqrt{\varepsilon_{0}^{2} + 2U
\varepsilon_{0} + \phi({\bf k})|^{2} \pm 2 (\varepsilon_{0} + U)
|\phi({\bf k})|}. \label{6}
\end{eqnarray}
If we take $U$ to be zero in Eq.(\ref{6}), it is same as
Eq.(\ref{4}). Furthermore, for  $0 < \beta < 2$, the two bands
keep touching and the Dirac point structure is not destroyed by
the interaction, which is shown explicitly in the following.

For $0 < \beta < 2$, there are two regions which we are interested
in. The first one is around the ${\bf k} = 0$, by expanding the
momentum ${\bf k}$ around $(0,0)$ and up to the second order of
$k_{x}$ and $k_{y}$, the dispersion relation (\ref{6}) becomes
\begin{eqnarray}
\varepsilon_{-}({\bf k}) = \sqrt{v_{x}^{2} k_{x}^{2} + v_{y}^{2}
k_{y}^{2}}, \label{7}
\end{eqnarray}
where $v_{x} = \sqrt{3U \beta/2 (2 + \beta)}$ and $v_{y} =
\sqrt{U/2}$. The sound velocity is anisotropic and dependent on
the anisotropy that is characterized by $\beta$ of the lattice.
The second region is around the Dirac points. Following the same
procedure, we obtain
\begin{eqnarray}
\varepsilon_{\pm}({\bf q})&=&\varepsilon_{0}\pm\sqrt{\tilde{v}_{x}^{2}q_{x}^{2}+\tilde{v}_{y}^{2}q_{y}^{2}}, \label{8}\\
\varepsilon_{\pm}({\bf q})&=&\sqrt{\varepsilon_{0}^{2}+2U\varepsilon_{0}}\pm\frac{\varepsilon_{0}+U}{\sqrt{\varepsilon_{0}^{2}+2U\varepsilon_{0}}}
\sqrt{\tilde{v}_{x}^{2}q_{x}^{2}+\tilde{v}_{y}^{2}q_{y}^{2}}, \nonumber \\
\label{9}
\end{eqnarray}
where $\bf q$ is a small momentum away from the Dirac points
${\bf K}=(k_{x}^{0},k_{y}^{0})$, i.e. $(k_{x},k_{y})=(k_{x}^{0}
+q_{x},k_{y}^{0}+q_{y})$. $\tilde{v}_{x}=\sqrt{3}\beta/2$ and
$\tilde{v}_{y}=\sqrt{1-\beta^{2}/4}$. The anisotropy of the velocity
is a consequence of the anisotropy of the lattice. Removing the
constant energy, we see that $\varepsilon_{\pm}({\bf q})$ represents
the standard energy-momentum relation for the relativistic Dirac
particles, with $v_{x}$ and $v_{y}$ replacing the light velocity.
Therefore, there is real Dirac structure around the touching
points. Furthermore, compared Eq.(\ref{9}) to Eq.(\ref{8}), we see
that the interaction has only modified the velocity of quasiparticles
around the Dirac points.

Before discussing how to detect the Dirac point structure in
experiments, we have to realize that here the physics is in the
ultracold and dilute region, the quantum depletion is very small,
e.g. for  $^{23}$Na at a typical density of $10^{14}$ cm$^{3}$,
the quantum depletion is $0.2\%$. The quantum depletion for two
dimensional system is even more smaller\cite{K. Xu}. Therefore,
unlike other cases, here the physics is dominated by the
condensates and low energy excitation. As a result, the effects of
the Dirac points are very weak for a ${\bf k}=0$ condensate to be
detected. This is the reason why no signatures of Dirac points is
observed in experiments\cite{Parvis Soltan-Panahi1,Parvis
Soltan-Panahi2}. To observe the effects of the Dirac points, we
have to drive the bosons to condense at ${\bf k} = {\bf K}$.
However, when the bosons are driven to condense at ${\bf k} = {\bf
K}$, there is modulational instability, this can be confirmed by
the discrete nonlinear Schr\"{o}dinger equation (DNLS)\cite{A.
Smerzi},
\begin{eqnarray}
i\hbar\frac{\partial\psi_{A}}{\partial t}&=&\sum_{s}t_{s}\psi_{Bs}+U|\psi_{A}|^{2}\psi_{A}, \nonumber \\
i\hbar\frac{\partial\psi_{B}}{\partial t}&=&\sum_{\bar{s}}t_{\bar{s}}\psi_{A\bar{s}}+U|\psi_{B}|^{2}\psi_{B}, \label{15}
\end{eqnarray}
where $s(\bar{s})$ denotes the nearest-neighbor vectors from A(B)
to B(A). The stationary solutions of Eq.(\ref{15}) are plane waves
$\psi_{0}^{A,B} exp[i(\bf{k}\cdot \bf{A}(\bf{B})-\nu t)]$, of
frequency $\nu=\pm|\phi({\bf k})|+U$ (here we assume $n_{0}^{A} =
|\psi_{0}^{A}|^{2} = n_{0}^{B} = |\psi_{0}^{B}|^{2} = n_{0}=1$).
To check the stability of these states, we perturb the carrier
wave with small amplitude phonons: $\psi_{A,B} = (\psi_{0}^{A,B} +
u_{A,B} e^{i\bf{q} \cdot \bf{A} (\bf{B})} + v^{*}_{A,B}
e^{-i\bf{q} \cdot \bf{A}(\bf{B})}) e^{i(\bf{k} \cdot \bf{A}
(\bf{B}) - \nu t)}$, The DNLS excitation spectrum is determined by
\begin{eqnarray}
&&\lambda^{4}-[2\xi_{{\bf k}}^{2}+4U\xi_{{\bf k}}+|\phi({\bf k+q})|^{2}+|\phi({\bf k-q})|^{2}]\lambda^{2} \nonumber \\
&&-2(U+\xi_{{\bf k}})[|\phi({\bf k+q})|^{2}-|\phi({\bf k-q})|^{2}]\lambda+(\xi_{{\bf k}}^{2}+2U\xi_{{\bf k}})^{2} \nonumber \\
&&-U^{2}|\phi({\bf k+q})^{*}+\phi({\bf k-q})|^{2}+|\phi({\bf k+q})|^{2}|\phi({\bf k-q})|^{2} \nonumber \\
&&-(\xi_{{\bf k}}^{2}+2U\xi_{{\bf k}})[|\phi({\bf k+q})|^{2}+|\phi({\bf k-q})|^{2}]=0, \label{16}
\end{eqnarray}
where $\xi_{{\bf k}} = |\phi({\bf k})|$. When ${\bf k} = 0$, the
spectrum is given as $\lambda_{\pm} = \sqrt{\varepsilon_{0}^{2} +
2U \varepsilon_{0} + |\phi({\bf k})|^{2} \pm 2(\varepsilon_{0} +
U) |\phi({\bf k})|}$, the same as Eq.(\ref{6}) and therefore it is
just the Bogoliubov spectrum. When ${\bf k} = {\bf K}$, imaginary
will naturally appear in the excitation spectrum, and as a result,
the condensate is modulationally unstable.

In fact, when the bosons are condensed at $\bf k=\bf K$, we can even see that
the Dirac points are destroyed by the interaction within the Bogoliubov theory.
Based on the Bogoliubov theory, the Hamiltonian is given as
$H_{BG}=\frac{1}{2}\sum_{q}B_{\bf q}^{\dag}H_{\bf q}B_{\bf q}$,
where $B_{\bf q}^{\dag}=(a_{\bf q}^{\dag},a_{-\bf q},b_{\bf q}^{\dag},b_{-\bf q})$ and
\begin{eqnarray}
H_{{\bf q}}=\left(\begin{array}{cccc} U & U& \phi(\bf q+K) & 0 \\
U & U& 0 & \phi(\bf q+K) \\
\phi^{*}(\bf q+K) & 0  & U & U  \\
0 & \phi^{*}(\bf q+K) & U   & U \end{array}\right). \nonumber
\end{eqnarray}
The energy spectrum is given as
\begin{eqnarray}
\varepsilon_{\rm \scriptscriptstyle {\bf K}}({\bf q})=\sqrt{(|\phi({\bf q+K})|\pm2U)|\phi({\bf q+K})|}.\label{17}
\end{eqnarray}
We see no matter how weak the interaction is, imaginary always
appears in the long-wavelength limit, i.e. ${\bf q}\rightarrow0$.
As a result, the topology of the Dirac points will be altered by
the long-wavelength excitations, which agrees with the result
obtained in Refs.\cite{Zhu Chen, O. B. Treidel}. All these results
indicate that to observe the Dirac point structure in Bose
condensates is a challenge. Recently, a way to handle this problem
of stability was introduced\cite{L. H. Haddad3} and the vortex
patterns suggested there can be used as a probe of the Dirac point
structure in experiments.

The study above is based on the tight-binding model. We know the
tight-binding model is based on the assumption that the lattice is
deep enough to guarantee that the Wannier wave function is
well-localized. However, for a lattice that is not sufficiently
deep, the tight-binding model with just the NN hopping is much
less accurate than the one with both NN hopping and the NNN
hopping\cite{J. I. Azpiroz}, and therefore, the NNN hopping should
be taken into account for such not sufficiently deep lattice. In
the following, we will consider the NNN hopping to see what
effects the NNN hopping has to the condensate.

\begin{figure}[bp]
\subfigure{\includegraphics[width=4.5cm,height=4cm]{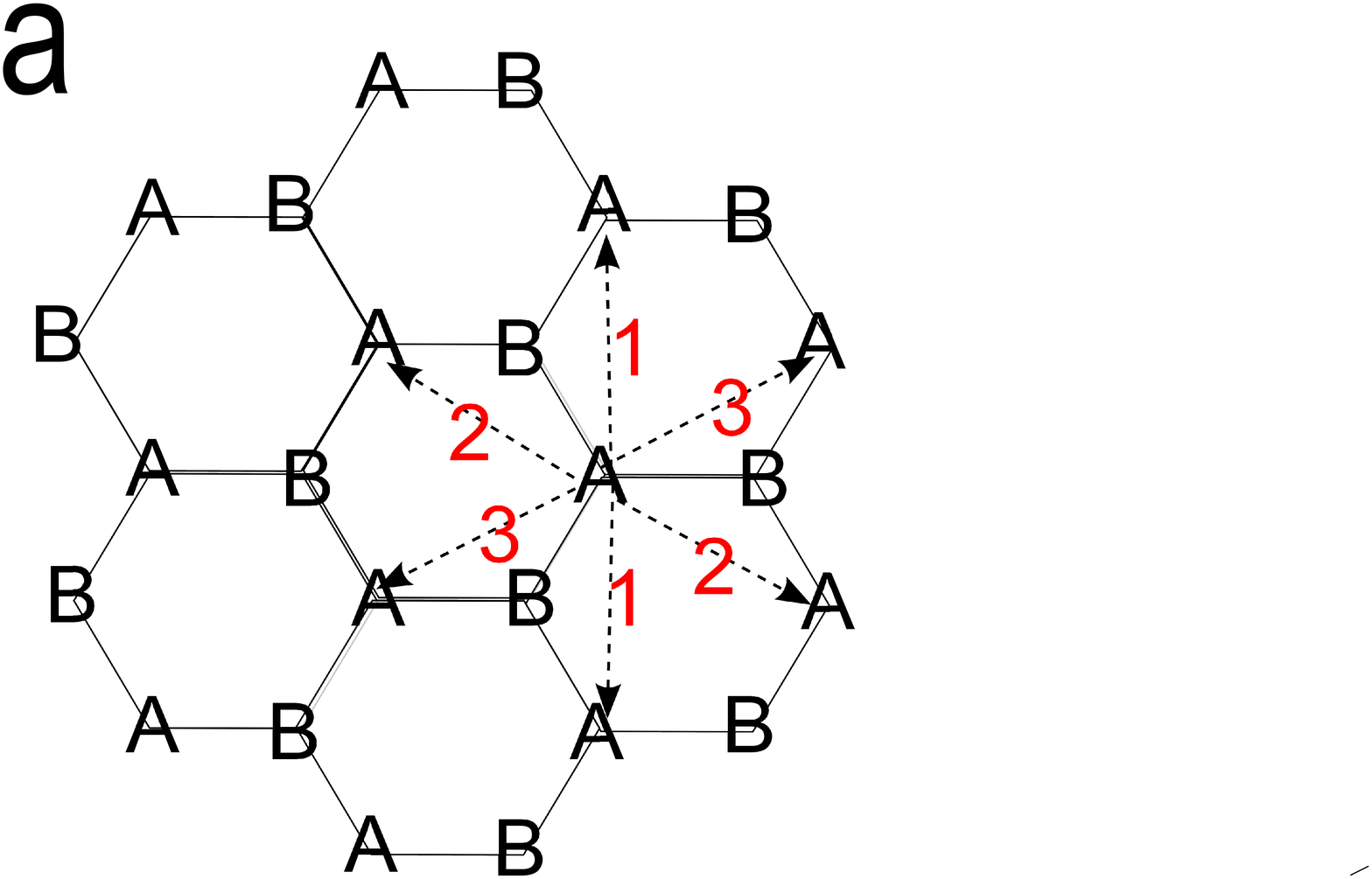}}
\label{Fig.3a}
\subfigure{\includegraphics[width=4cm,height=4cm]{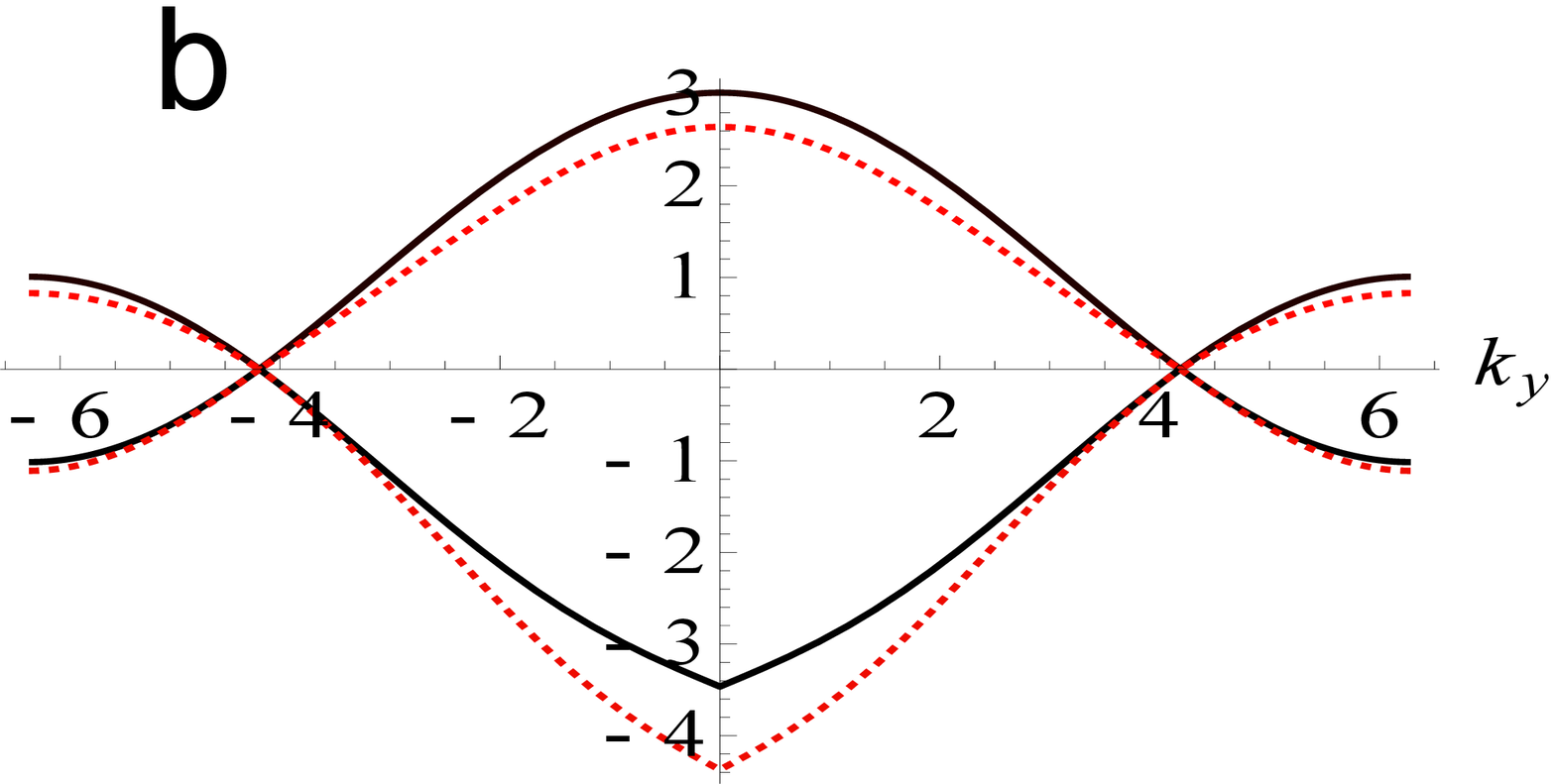}}
\label{Fig.3b} \caption{(Color online) (a) 1,2,3 stand for three
different kind of hopping. (b) The common parameters for both the
solid
 line (black) and the dashed line (red) are $\beta=\lambda=1$,  $U=0.5$ and $k_{x}=0$, while $\gamma=0$ for the solid line
 and  $\gamma=0.1$ for the dashed line.  \label{Fig.2}}
\end{figure}

\begin{figure}[bp]
\subfigure{\includegraphics[width=4cm,height=3cm]{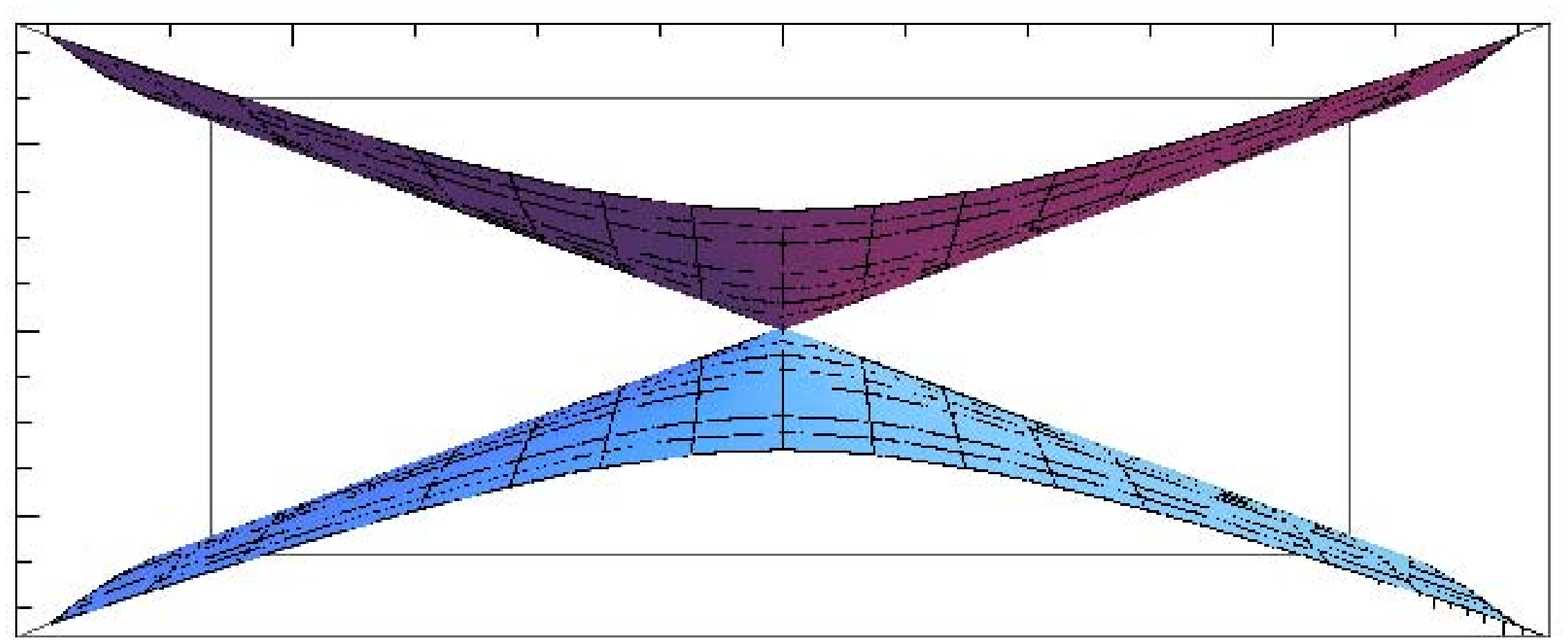}}
\subfigure{\includegraphics[width=4cm,height=3cm]{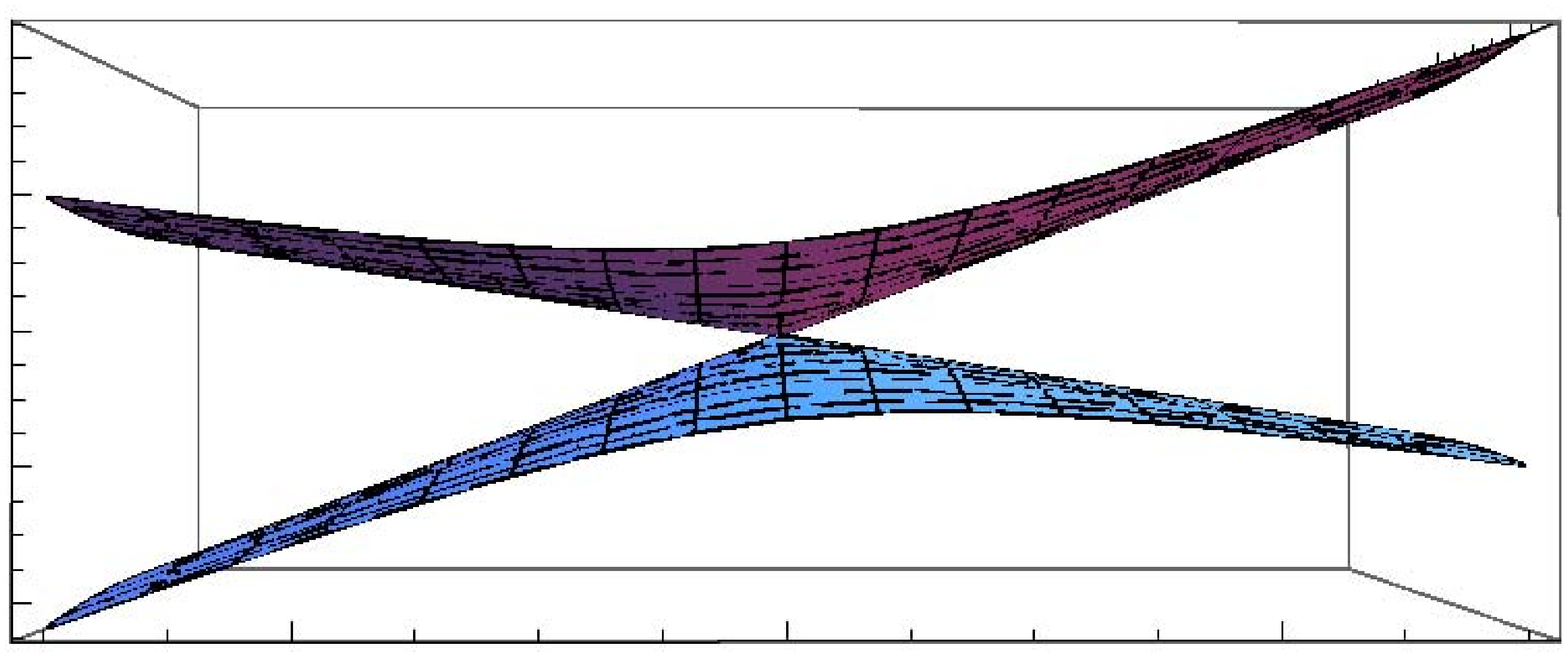}}
\caption{(Color online) (Right) The isotropic case, i.e.
$\beta=\lambda=1$; (Left) The anisotropic case, i.e.
$\beta\neq\lambda$. \label{Fig.3}}
\end{figure}

After we introduce the NNN hopping into Eq.(\ref{1}), the excitation spectrums take the form
\begin{eqnarray}
\tilde{\varepsilon}_{\pm}({\bf k})=-\tilde{\mu}-h({\bf
k})\pm|\phi({\bf k})|, \label{10}
\end{eqnarray}
where
\begin{eqnarray}
\tilde{\mu}&=&-2\gamma(1+2\lambda)-\beta-2, \nonumber \\
h({\bf k})&=&2\gamma(\cos k_{y}+2\lambda\cos\frac{\sqrt{3}k_{x}}{2}\cos \frac{k_{y}}{2}), \nonumber
\end{eqnarray}
where $\gamma$ is the NNN hopping constant in unit of $t$ and
$\lambda$ is the anisotropic of the NNN hopping. Based on the
anisotropy of the lattice and the anisotropy of the NN hopping,
the hopping amplitudes along the path labelled by $2$ and $3$ in
Fig.\ref{Fig.2}(a) are equal but different to the one along the
path labelled by $1$. From Eq.(\ref{10}), it is direct to see that
the two bands still touch. However, when we expand the dispersion
relation around the Dirac points and keep only the terms related
to the lowest order of momentum, we find
\begin{eqnarray}
\tilde{\varepsilon}_{\pm}({\bf q})=-2\gamma(\beta-\lambda)
\sqrt{1-\frac{\beta^{2}}{4}}q_{y}\pm\sqrt{\tilde{v}_{x}^{2}q_{x}^{2}+\tilde{v}_{y}^{2}q_{y}^{2}},
\label{11}
\end{eqnarray}
this indicates that if the anisotropy of the NNN hopping is
different from the NN hopping, the Dirac point structure is no
longer well defined even though the two bands still touch with
each other (as shown in Fig.\ref{Fig.3}). In the harmonic
approximation, we calculate the NN hopping and the NNN hopping and
find when $\beta\neq1$, $\lambda$ is not equal to $\beta$ (if we
further consider the next-next-nearest-neighbor (NNNN) hopping,
the NNNN hopping will induce a small gap for the case of
$\beta\neq1$). Therefore, to observe the Dirac point structure,
it's better to choose the system to be isotropic, i.e. $\beta =
\lambda = 1$. In the following, we set $\beta = \lambda = 1$.

When the interaction is tuned on, the dispersion relations change into
\begin{eqnarray}
\tilde{\varepsilon}_{\pm}({\bf k})=\sqrt{\epsilon_{{\bf
k}}^{2}+2U\epsilon_{{\bf k}} +|\phi({\bf
k})|^{2}\pm2(\epsilon_{{\bf k}}+U)|\phi({\bf k})|}, \label{12}
\end{eqnarray}
where $\epsilon_{{\bf k}}=-\tilde{\mu}-h({\bf k})$. By expanding
the dispersion relation around $k=0$, we obtain
\begin{eqnarray}
\tilde{\varepsilon}_{-}({\bf k})=\sqrt{\bar{v}_{x}^{2}k_{x}^{2}+\bar{v}_{y}^{2}k_{y}^{2}}, \label{13}
\end{eqnarray}
where the modified velocity $\bar{v}_{x}=\bar{v}_{y}=\sqrt{\frac{U}{2}(1+6\gamma)}$. Compared to
$v_{x}=v_{y}=\sqrt{U/2}$ in Eq.(\ref{7}) with $\beta=1$, we see the sound velocity is  effectively
increased due to the NNN hopping (as illustrated in Fig.\ref{Fig.2}(b)). By expanding the dispersion
relation around the Dirac points, we obtain,
\begin{eqnarray}
\tilde{\varepsilon}_{\pm}({\bf q})=A(\gamma,U)\pm\frac{B(\gamma,U)}{\sqrt{A(\gamma,U)}}
\sqrt{\tilde{v}_{x}^{2}q_{x}^{2}+\tilde{v}_{y}^{2}q_{y}^{2}}, \label{14}
\end{eqnarray}
where $A(\gamma,U)=9(1+\gamma)^{2}+6U(1+\gamma)$ and
$B(\gamma,U)=3(1+\gamma)+U$. By a direct calculation with concrete
parameters, we find that unlike the case around ${\bf k}=0$, the
NNN hopping nearly does not affect the velocity around the Dirac
points (also shown in Fig.\ref{Fig.2}(b)). Based on these results,
we obtain the conclusion that for an isotropic lattice, the NNN
hopping mainly affects the dynamics of the modes around the ${\bf
k}=0$, it nearly does not affect the dynamics of the modes around
the Dirac points; However, for an anisotropic lattice, the NNN
hopping not only affects the dynamics of the modes around the
${\bf k}=0$, but also alters the topology of the Dirac points and
makes the Dirac points no longer well defined.

From the above analysis, we see that to guarantee that the
tight-binding model gives an accurate description to the
condensates, the lattice indeed has to be sufficiently deep to
guarantee that the NNN hopping is far smaller than the NN hopping,
i.e. $\gamma \ll 1$.

In summary, we have studied the Bose-Einstein condensates in a
honeycomb lattice and found that for a ${\bf k} = 0$ condensate,
the Dirac points appear in the Bogoliubov excitation spectrum when
$0 < \beta < 2$, and when the bosons are condensed at ${\bf k} =
{\bf K}$, the topology of the Dirac points will be altered by
arbitrary weak interaction. Furthermore, we find that the NNN
hopping of not small strength has obvious effects to the dynamics
of the condensates, and therefore it should be taken into account
when the lattice is not sufficiently deep.

This work is supported by NSFC Grant No.11275180.

\end{document}